# Cooper-Pair Molasses - Cooling a Nanomechanical Resonator with Quantum Back-Action

A. Naik[1,2], O. Buu[1,3], M.D. LaHaye[1,3], A.D. Armour[4], A. A. Clerk[5], M.P. Blencowe[6], K.C. Schwab[1]

[1]*Laboratory for Physical Sciences, 8050 Greenmead Drive, College Park, MD, 20740 USA*

[2]*Department of Electrical and Computer Engineering, University of Maryland, College Park, MD 20740 USA*

[3]*Department of Physics, University of Maryland, College Park, MD 20740 USA*

[4]*School of Physics and Astronomy, University of Nottingham, Nottingham, NG7 2RD, United Kingdom*

[5]*Department of Physics, McGill University, Montreal, QC Canada H3A 2T8*

[6]*Department of Physics and Astronomy, Dartmouth College, Hanover, NH 03755 USA*

**Quantum mechanics demands that measurement must affect the measured object. When continuously monitoring position with a linear amplifier the Heisenberg uncertainty relationship requires that the object be driven by force impulses, called back-action[1,2,3]. Here we report the detection of the measurement back-action of a superconducting single-electron transistor (SSET) which measures the position of a radio-frequency nanomechanical resonator. The SSET exhibits non-trivial quantum noise properties, with the back-action manifesting itself as an effective thermal bath which depends sensitively on the SSET bias point. Surprisingly, when biasing near a transport resonance, we observe cooling of the**

**nanomechanical mode from 550 mK to 300 mK. These measurements have implications for nanomechanical readout of quantum information devices and the limits of ultra-sensitive force microscopy, e.g. single nuclear spin magnetic resonance force microscopy. Furthermore, we anticipate the use of these back-action effects to prepare ultra-cold and quantum states of mechanical structures, which would not be accessible with existing technology.**

In practice, these back-action impulses arise from the quantized and stochastic nature of the fundamental particles utilized in the measuring device. For example, in high precision optical interferometers such as the LIGO gravitational wave detector[4] or in the single-spin force microscope[5], the position of a test mass is monitored by reflecting laser-light off of the measured object and interfering this light with a reference beam at a detector. The measured signal is the arrival rate of photons, and one might say that the optical "conductance" of the interferometer is modulated by the position of the measured object. Back-action forces which stochastically drive the measured object result from the random impact and momentum transfer of the discrete photons. This mechanical effect of light is thought to provide the ultimate limit to the position and force sensitivity of an optical interferometer. Although this photon "ponderomotive" noise has not yet been detected during the measurement of a macroscopic object[6], these back-action effects are clearly observed and carefully utilized in the cooling of dilute atomic vapors to nano-Kelvin temperatures.

In the experiments reported here, we study an SSET which is capacitively coupled to a voltage-biased ($V_{NR}$), doubly-clamped nanomechanical resonator (Fig. 1). Like the interferometer, the conductance of the SSET is a very sensitive probe of the resonator's position, whereas the particles transported in this case are a mixture of single and Cooper-paired electrons. We have recently shown the SSET to be nearly a quantum-limited position detector[7], however reaching the best sensitivity will ultimately be limited by the back-action of the charged particles[3], which could not be observed in previous experiments because of insufficient SSET-resonator coupling.

The back-action force of the SSET results in three measurable effects on the resonator: a frequency shift, a damping rate, and position fluctuations. The frequency shift and damping rate are caused by the in-phase and small out-of phase response in the average electrostatic force between the SSET and resonator, as the resonator oscillates. Position fluctuations arise from fluctuations in this force. As electrons and Cooper-pairs are transported through the SSET island, the island charge changes stochastically by +/-1 or 2 electrons, causing the electrostatic force to fluctuate with amplitude of $10^{-13} N_{RMS}$, with a white spectral density of $\sqrt{S_F} \sim 10^{-18} \frac{N}{\sqrt{Hz}}$, extending to ~20 GHz (assuming $V_{NR}$ = 2V and parameters in the Supplementary Information (SI).)

These back-action effects are observed by measuring the SSET charge noise in the vicinity of the resonator frequency (see SI) thus obtaining the resonator noise temperature, $T_{NR}$, and dissipation rate, $\gamma_{NR} = \frac{\omega_{NR}}{Q_{NR}}$ for different coupling voltages, $V_{NR}$, and refrigerator temperatures, $T_{Bath}$. Figure 2 inset shows an example of the measured spectral noise power where the mechanical resonance at $\omega_{NR} = 2\pi \cdot 21.837$ MHz is clearly visible, and accurately fits a simple harmonic oscillator response function, on top of a white power spectrum due to an ultra-low noise microwave preamplifier used to read out the SSET with microwave reflectometry[8].

For low SSET-nanoresonator coupling strengths, and the SSET biased close to the Josephson Quasiparticle Peak (JQP)[9], $T_{NR}$ simply follows $T_{Bath}$, with the lowest temperature point corresponding to occupation number $N_{TH}$~25, and the nanoresonator shows a very low dissipation rate ($Q_{NR}$ = 120,000 at $T_{Bath}$ = 100mK); see Figure 2. This is an elementary demonstration of the Equipartition Theorem and of nanomechanical noise thermometry[7].

As the SSET bias point is held fixed at JQP and the coupling voltage is increased, we find clear signatures of all three expected back-action effects. Figure 3 shows the change in resonance frequency and dissipation rate, $\gamma_{NR}$, caused by coupling to the SSET. The back-action fluctuation effects on $T_{NR}$ first become noticeable for $T_{Bath}$ <200mK (Figure 2), where we observe an increase and ultimately a saturation of $T_{NR}$~200mK; the SSET back-action exerts a stochastic force drive, in excess of the thermal noise of $T_{Bath}$.

At our highest coupling strength, $V_{NR}$ = 15V, we find $T_{NR}$ has a much weaker dependence on $T_{Bath}$, tending toward a fixed value of ~200mK. In this high coupling limit and for $T_{Bath}$ > 200mK, we observe the counter-intuitive effect that a noisy, non-equilibrium measuring device can cool the nanomechanical mode.

This behavior can be understood through models based on a quantum noise approach[10] or a master equation description of the coupled electro-mechanical system[11]. These models show that the back-action effects are that of an effective thermal bath; similar results have been obtained for oscillators coupled to other non-equilibrium devices: a normal state SET[12,13], a quantum point contact[14], and in experiments with a ping-pong ball in turbulent air flow[15]. The resulting damping and temperature of the resonator due to coupling to both refrigerator thermal bath ($T_{Bath}$) and SSET effective bath ($T_{SSET}$) through the associated damping rates ($\gamma_{Bath}$ and $\gamma_{SSET}$) are then given by:

$$\gamma_{NR} = \gamma_{Bath} + \gamma_{SSET} \tag{1a}$$

$$\gamma_{NR} T_{NR} = \gamma_{Bath} T_{Bath} + \gamma_{SSET} T_{SSET} \tag{1b}$$

As long as $\gamma_{NR}$ is positive, equations 1a-b provide a simple interpretation of the data in Figure 2: as the coupling to the SSET is increased, the damping $\gamma_{SSET}$ increases, so the effective temperature of the resonator is pulled away from $T_{Bath}$ and closer to $T_{SSET}$. Experimentally, $T_{SSET}$ is easily determined from the value of $T_{Bath}$ where all the curves cross, or the value of $T_{NR}$ when $T_{Bath} \to 0$: we find $T_{SSET} \sim 200$mK. The curvature observed in $T_{NR}$ at high couplings is likely due to the temperature dependence of the $\gamma_{Bath}$. Note that $T_{SSET}$ is not determined by the thermodynamic temperature of the carriers in the SSET, but is a measure of the intensity of the charge fluctuations on the SSET.

Using the quantum noise approach[3,16], one finds that the back-action effects are determined by the asymmetric-in-frequency quantum noise spectrum of the back-action force, $S_F(\omega)$:

$$2m \cdot \hbar \omega_{NR} \cdot \gamma_{SSET} = S_F(\omega_{NR}) - S_F(-\omega_{NR}) \tag{2a}$$

$$2m \cdot k_B T_{SSET} \cdot \gamma_{SSET} = S_F(\omega_{NR}) + S_F(-\omega_{NR}) \tag{2b}$$

where $k_B$ is Boltzmann's constant and $m$ is the oscillator mass, and the positive (negative) frequency noise power describes the noise responsible for the rates of energy emission (absorption) from the nanoresonator to the SSET. Thus our nanomechanical resonator provides a frequency-resolved measurement of the asymmetric, quantum noise properties of the SSET[17,18,19], albeit only at the resonator frequency.

When biased at JQP, the transport is resonant and involves a process of Josephson and quasi-particle tunneling. This leads to a simple expression for $T_{SSET}$:

$$k_B T_{SSET} = \frac{\hbar}{4} \frac{\Gamma^2 + 4(\Delta E/\hbar)^2}{4\Delta E/\hbar} \qquad (3)$$

where, $\Delta E$ is the difference between final and initial energies of the tunneling Cooper-pair, and $\Gamma$ is the quasiparticle tunnel rate, which is essentially temperature independent far below the superconducting transition temperature. This result is analogous to laser ponderomotive cooling of a mechanical cavity[20], or to Doppler cooling of a two-level atom[21]; hence we call this "Cooper-pair molasses."

Figure 4 shows the sensitive dependence of $T_{SSET}$, $\gamma_{SSET}$, and $\omega_{NR}$, as we tune the SSET bias point through the two neighboring JQP resonances. Between the two JQP peaks where $\Delta E>0$ ("red detuned"), $T_{SSET}$ is in excellent agreement with the theoretical predictions (~220mK) (Fig. 4d). The changes in frequency (Fig. 4b) are also reasonably well described, indicating the overall magnitude of the non-fluctuating component of the back-action force is close to expectation. Only qualitative agreement is achieved for the damping, which results from a very small asymmetry in the force noise power (Eq. 2a). We observe the expected quadratic dependence on $V_{NR}$ (Fig. 3) and the increase in damping as one approaches the JQP resonance (Fig. 4c). However, the magnitude is 14 times larger than expected. We believe this excess may be related to the longstanding discrepancy between theory and experiment in transport characteristics at the JQP[22]. We do not believe that the excess damping observed is due to another independent bath as the dissipation is clearly very sensitively controlled by the SSET bias point (Figure 4.)

Outside the JQP resonances, the tunneling Cooper pairs can emit energy to the resonator ($\Delta E<0$, "blue-detuned"), and the linear theory predicts negative SSET temperatures and damping rates[10,11]. When the predicted $\gamma_{NR}$ becomes negative (region shaded blue in Fig. 4), states of small amplitude of motion are unstable with respect to a state of large amplitude motion of the resonator, where the amplitude is ultimately limited by non-linearity in the dynamics[10], and the picture of the SSET as an effective thermal bath breaks down. A theoretical description of our system in these parameter regions is beyond the scope of this paper, but we do find experimentally points of dramatically increased $T_{NR}$ outside the JQP peaks, as one would expect qualitatively. Additionally, we have observed the unstable behavior when biasing the SSET at the Double Josephson Quasiparticle Peak where the back-action effects are predicted to be larger than at JQP[10,11]. Although we are only able to make quantitative comparison for the regions of stable bias, the fact that SSET operating points with opposite sign $\Delta E$ yield the same current, but show very different back-action, provides compelling evidence that the SSET is the controlling source of the back-action fluctuations.

We can use Equation 2 to calculate the amplitude of the back-action impulses and compare this to the uncertainty principle requirement, $(S_x S_F)^{1/2} \geq \frac{\hbar}{2}$, where $S_x$ is the spectral power density of position (at our highest coupling voltage, we achieve $\sqrt{S_x} = 3 \cdot 10^{-16} \frac{m}{\sqrt{Hz}}$). Using $S_x=S_I/(dI/dx)^2$, and $S_I=2eI$ at the JQP resonance[23], we find $\sqrt{S_F S_x} = 15\frac{\hbar}{2}$, a factor of 15 larger than the uncertainty principle requirement. This excess back-action will limit the position resolution $\Delta x$ to a factor of 3.9 over the quantum limit (see SI). However, this level of ideality is already sufficient for an SSET to enable the realization of novel quantum states of the mechanical resonator including squeezed[24] and entangled states[25].

The nanomechanical cooling and heating effects discovered here are closely related to proposed processes using a Cooper-pair qubit[26,27] and a quantum dot [28,29], and are the first example of the interaction between coherent electronics and nanomechanical modes. Although the cooling power of our Cooper-Pair Molasses is fantastically small, $\dot{Q} \sim \gamma_{SSET} k_B (T_{Bath} - T_{SSET}) = 10^{-21} W$, it is sufficient to cool a single, well isolated, mechanical degree of freedom. Furthermore, we believe it is possible to use this method to produce ultra-cold, quantum states in nanomechanical systems with high resistance tunnel junctions since it is expected[10] that $T_{SSET} \propto R_J^{-1}$ and $\gamma_{SSET} \propto R_J^2$, providing increasing coupling to a decreasing temperature bath as the junction resistance, $R_J$, is increased. For instance, the junctions reported by Nakamura et al.[30] should reach a minimum $T_{SSET}$=0.3mK, providing access to the quantum ground state of a resonator similar to that reported here.

**Author Contributions**

A. Naik and O. Buu contributed equally to this work.

**Acknowledgement**

We thank A. Rimberg and A. Vandaley for helpful discussions and B. Camarota for assistance with the fabrication of the samples. M.P.B. is supported by the NSF under NIRT grant CMS-0404031. A.D.A. is supported by the EPSRC under grant GR/S42415/01. A.A.C. is supported by NSERC under grant RPGIN 311856.

# Figure Captions

**Figure 1 | Nanodevice and measurement schematic.** A colorized SEM micrograph of the device: a 21.9 MHz, doubly-clamped, SiN and Al nanomechanical resonator (NR) coupled to a SSET, with simplified measurement circuit schematics. See SI for sample parameters.

**Figure 2 | Resonator temperature versus bath temperature and coupling voltage.** Plot shows the effective temperature of the resonator (in mK and quanta $N_{TH}$), $T_{NR}$, versus $T_{Bath}$ with the SSET biased near the JQP. See text for discussion. Inset shows an example of the measured charge noise power taken at $T_{Bath}$ = 100 mK and $V_{NR}$ = 4V shown as red points and Lorentzian fit shown as black line. The area of this peak is proportional to the resonator temperature, $T_{NR}$.

**Figure 3 | Resonator damping rate and frequency shift versus coupling voltage.** shows $\gamma_{NR}$ versus $T_{Bath}$ and $V_{NR}$. The inset shows the resonant frequency versus $V_{NR}^2$ ($T_{Bath}$=30mK.) Both the frequency shift and the $\gamma_{SSET}$, scale as $V_{NR}^2$, as expected. At low $V_{NR}$, $\gamma_{NR}$ is asymptotic to $\gamma_{Bath}$. At high $V_{NR}$, $\gamma_{NR}$ is asymptotic to a quadratic dependence, $\gamma_{SSET}$, although at a rate 14 times higher than expected ($\gamma_{NR}$ --theory.) The figure above shows the nanomechanical mode coupled to the thermal baths, $T_{Bath}$ and $T_{SSET}$, through dissipative links, $\gamma_{Bath}$ and $\gamma_{SSET}$. The red (blue) arrows indicate the emission (absorption) of quanta by the baths leading to heating (cooling) of the resonator.

**Figure 4 | Back-action effects versus SSET bias point.** Comparison of experimental measurements (black points and lines) and theoretical predictions (blue and red lines) as $V_G$ is scanned through neighboring JQP resonances, with fixed $V_{DS}$ = 4$E_C$ - 100μV, and $V_{NR}$=5V. The panels show (**a**) the SSET current, (**b**) the resonator frequency shift, (**c**) the

resonator damping rate, $\gamma_{NR}$ and (**d**) the effective resonator temperature, $T_{NR}$. Note that in panel (**c**) the theoretically calculated values have been multiplied by a factor of 14. In (**c**) and (**d**), expected unstable bias points, $\{T_{NR}, \gamma_{NR}\}<0$, are shown in blue where SSET is "blue detuned", and stable bias points, $\{T_{NR}, \gamma_{NR}\}>0$ are shown in red where SSET is "red detuned".

*Figure #1*

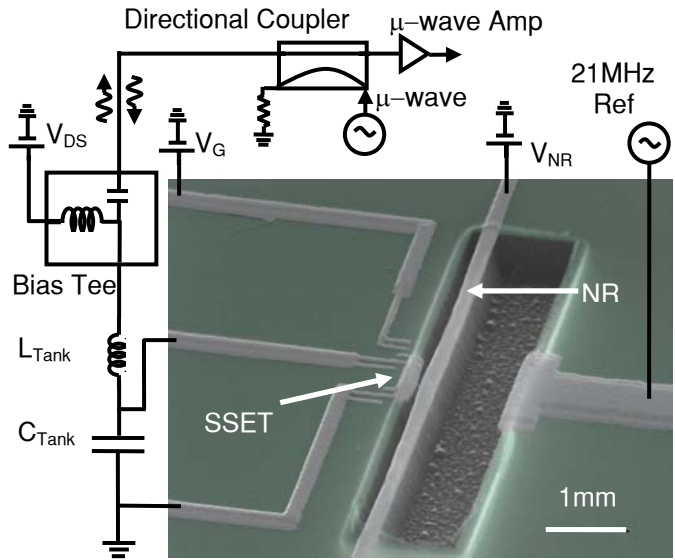

*Figure #2*

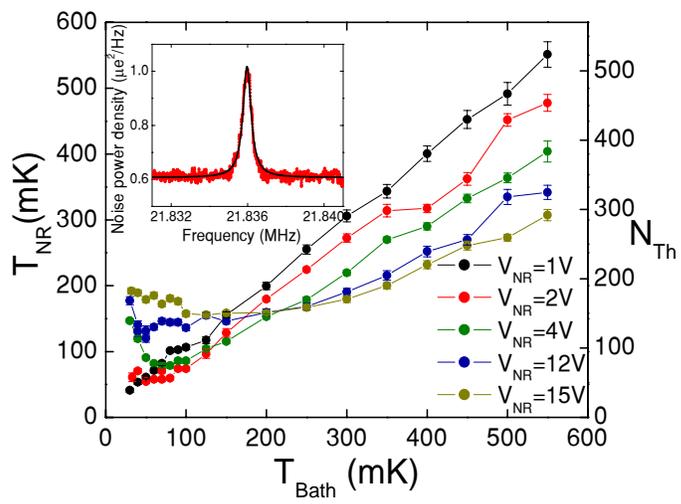

*Figure #3*

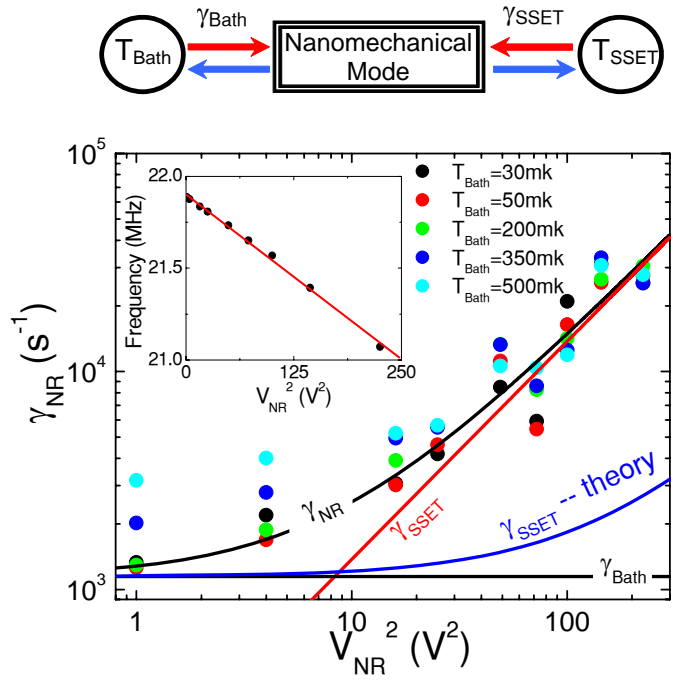

*Figure #4*

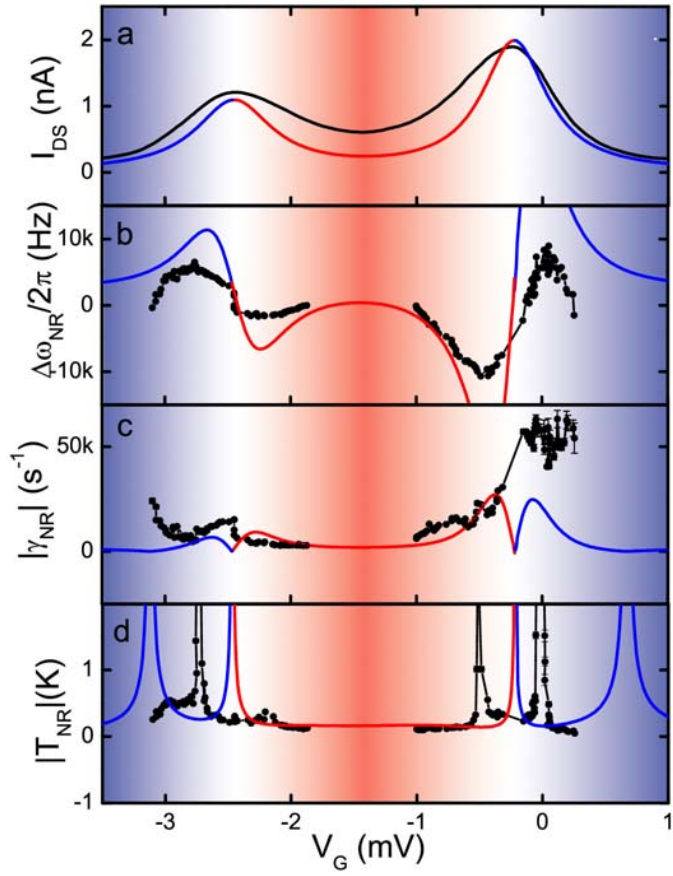

**Supplementary Information**

**"Cooper-Pair Molasses: Cooling a Nanomechanical Resonator with Quantum Back-action"**

A. Naik, O. Buu, M.D. LaHaye, A.D. Armour, A. A. Clerk, M.P. Blencowe, K.C. Schwab

**Sample Fabrication**

The sample is fabricated on a p-type, 10 Ohm-cm, (1,0,0), silicon substrate, coated with 50 nm of low-stress, amorphous silicon nitride (SiN). The doubly-clamped, nanomechanical resonator is 8.7 μm long, and 200 nm wide, composed of 50nm of SiN and 90 nm of Al. The SSET island is 1 μm long, located about 100 nm away from the resonator. The tunnel junctions, made of AlOx, are approximately 70 nm X 60 nm.

An on-chip LC resonator is microfabricated for impedance matching the SSET to an ultra-low noise, 50 Ω cryogenic microwave amplifier (Berkshire Technologies Model # L-1.1-30H) with $T_N$=2K. The LC resonator is formed by an interdigitated capacitor and a planar Al coil. Our circuit demonstrates a resonance at 1.17 GHz with a quality factor of about 10. The measurement set-up is similar to that described in Ref.1.

**Sample Characteristics**

The list below details the sample characteristics for the device shown in Fig. 1.

| | | |
|---|---|---|
| $C_1$ | =181 aF +-9aF | Junction capacitance of junction 1 |
| $C_2$ | =199 aF +-20aF | Junction capacitance of junction 2 |
| $C_{G\,NR}$ | =10.7aF +- 0.1aF | Capacitance between SSET and NR gate |
| $C_{NR}$ | =33.6aF +- 1aF | Capacitance between SSET and resonator |
| $C_G$ | =22.6aF+- 0.6 aF | Capacitance between SSET and SSET gate |
| $C_\Sigma$ | =449aF +-30aF | Total device capacitance |
| $R_\Sigma$ | =104kΩ +-2kΩ | Total device Resistance |
| $E_C$ | = 175 μV +-4μV | Coulomb blockade energy |
| $\Delta$ | = 192.0 μV+-0.7μV | Superconducting energy gap |
| $E_{J1}$ | = 13.0 μV | Josephson Energy for junction1 |
| $\Gamma_{a2}$ | = 67.4μV | 1st Quasiparticle tunneling rate through junction 2 |
| $\Gamma_{b2}$ | = 32.3μV | 2nd Quasiparticle tunneling rate through junction 2 |
| $E_{J2}$ | = 17.4 μV | Josephson Energy for junction 2 |
| $\Gamma_{a1}$ | = 50.4μV | 1st Quasiparticle tunneling rate through junction 1 |
| $\Gamma_{b1}$ | = 24.12μV | 2nd Quasiparticle tunneling rate through junction 1 |
| $R_{j1}$ | = 59.5kΩ | Resistance of junction 1 |
| $R_{j2}$ | = 44.5 kΩ | Resistance of junction 2 |

$dC_{NR}/dx$ = 0.3e-9 F/m      Derivative of the coupling capacitance
$k$ = 10 N/m      Spring constant
$\Delta F$ = 1.05e-13 N      Coupling strength at $V_{NR}$=1V
$\omega_{NR}$ = $2\pi$ x 21.866 MHz      Resonator intrinsic frequency
$I_{DS}$ = 0.8nA      Approx value during measurements
$dI_{DS}/dV_G$ = 9.4e-7 A/V      From the slope near the bias point
$\frac{dI_{DS}}{dx} = 12.5 A/m$      at $V_{NR}$=1V
$T_{SSET}$ = 200mK      Approx. experimental value.
$Q_{SSET}$ = $10^6/V^2$      Approx. experimental value.
$Q_{Bath}$ = 120000      Intrinsic quality factor (measured at 30mK 1V)

In the notation above the junction with capacitance $C_1$ is at high voltage while $C_2$ is at the ground potential. The quantities $C_{G\ NR}$, $C_{NR}$ and $C_G$ are obtained from the periodicity of the current modulation curves; $C_1$ and $C_2$ are calculated using the slopes of the resonances in the $I_{DS}$-$V_{DS}$-$V_G$ map (see figure S1), $R_\Sigma$ is determined from the $I_{DS}$-$V_{DS}$ curves at large drain-source bias $V_{DS} \gg 4\Delta$. The charging energy is measured from the position of the DJQP feature and JQP crossing. It is in good agreement with the value be calculated from the sum of all the capacitances, $e^2/(2C_\Sigma)$. The superconducting energy gap is calculated using the position of onset of quasiparticle current occurring at $4\Delta$ (figure S1).

The Josephson energy for each junction is given by $E_J = (R_Q \Delta / R_j\ 8)\ F\ (E_C/\Delta)$, where the function F(x) describes the renormalization of $E_J$ over the usual Ambegaokar-Baratoff value due to the finite value of $E_C$. In physical terms, the charging energy lowers the energy of the virtual state involved in a Josephson tunneling event, thus enhancing $E_J$; a detailed discussion of this effect and the analytic form of F(x) is given in Ref. [*2*]. Using this analytical form, we obtain F(x) = 1.26 for our device. The values of the two quasiparticle tunneling rates ($\Gamma$'s) for each resonance are calculated using the theoretical expressions (see for example Ref. [*3*]). The value of the individual junction resistances were extracted by comparing the experimentally measured ratio of peak currents for two adjacent JQP resonances (at the same $V_{DS}$) with the theoretical prediction (*3*). Figure 4 shows the theoretical prediction and the measured values of the current. The width of the measured JQP resonances is broader than that predicted by the theory. This discrepancy has been observed in other SSETs with $\Delta \sim E_c$ (*4*) and suggests that the current may contain contributions from other (presumably incoherent) processes beyond those associated with the JQP resonance.

The spring constant, $k$, is estimated from the effective mass of the resonator, $0.68 \times 10^{-15} kg$, which in turn estimated from the geometry of the beam, and renormalized by 0.99 to account for the shape of the first vibration mode.

The electromechanical coupling strength is defined by $\Delta F = 2 E_c \frac{V_{NR}}{e} \frac{dC_{NR}}{dx}$. The derivative $dC_{NR}/dx$ is obtained from 2-dimensional numerical calculations of the capacitances using FEMLAB. We trust the calculated value because the same simulations give $C_{NR}$=29aF, in good agreement with the experiment. Moreover, the value of the second derivative $d^2C_{NR}/dx^2$=0.004aF/(nm)$^2$ gives an electrostatic frequency shift $\frac{\Delta\omega}{\omega} = \frac{-V_{NR}^2}{2k}\frac{d^2C}{dx^2} = -2\times10^{-4}V_{NR}^2$, consistent with the measured value (1.6x10$^{-4}$/V$^2$). Note that this frequency shift is in addition to that arising from the SSET back-action near the JQP (*3*), but can be distinguished from it as it is independent of the SSET bias point.

Another consistency test is the amplitude of the thermal noise signal. In the absence of back-action, the integrated charge noise $<Q^2>$ induced on the SSET by the thermal motion of the resonator is proportional to the bath temperature, with a slope

$$\frac{d<Q^2>}{dT_{Bath}} = \frac{\Delta F^2}{4kE_c}.$$

The value of this slope, 9.8e-9$e^2$/mK, calculated from the above-determined parameters, compares favorably with 7.8e-9$e^2$/mK, determined from a linear fit of the data at $V_{NR}=1V$.

All the calculations for the paper are based on equations given in reference (*4*).

**Experimental method:**

For the measurements shown in Fig. 2 and 3, the device was biased at the point indicated by the red ellipse on Fig. S1. The precise location of the bias point is:

$V_{DS}$=(4-0.57) $E_c$      $V_G$=0.078 $e$ from the resonance      $I_{DS}$ =0.8nA

The SSET bias point is held fixed by monitoring the SSET current, $I_{DS}$, and applying a feedback voltage to a near-by gate electrode $V_G$. This allows us to counteract the low frequency charge noise which is typical in these devices. The thermal noise spectra of the resonator are recorded using a spectrum analyzer. A 20.5MHz charge signal is continuously applied to a nearby gate to monitor the charge sensitivity of the SSET in real time. The amplitude of the reference charge signal, 2me, was itself calibrated by using the Bessel-response technique (*5*).

Note that since we are using the radio-frequency SET technique to measure the thermal noise of the nanoresonator, we send a microwave excitation to $V_{DS}$ of the device. Because of this, our measurement has an "average" effect of constantly sweeping an elliptical area around the bias point. The major and minor axes of this ellipse are determined by the strength of the microwave and the reference charge signal at the SSET, which are, respectively 21µV and 15µV (peak-to-peak values). We estimate that this averaging effect could increase the amount of the damping due to SSET by a factor of ~2 as compared the value obtained at the center of the ellipse. The amplitudes of these signals are smaller than any of the features on the $I_{DS}$ map, approximating the ideal measurement with fixed bias point.

**Thermometry and Data Analysis:**

The charge induced at the SSET island from the voltage biased nanomechanical resonator is given by $Q_{SSET}=C_{NR}V_{NR}$, where $C_{NR}$ is the resonator-SSET capacitance. Motion of the resonator, will modulate the capacitance, $C_{NR}$, which will change the SSET charge by

$$\delta Q_{SSET} = \frac{dC_{NR}}{dx}V_{NR}\delta x$$

Thus mechanical noise will produce charge noise,

$$S_Q^{NR}(\omega) = \left(\frac{dC_{NR}}{dx}V_{NR}\right)^2 S_x(\omega)$$

where $S_Q^{NR}(\omega)$ and $S_x(\omega)$ are the charge and position noise power spectral densities. Thermal motion of the resonator is expected to have a spectral density given by:

$$S_x(\omega) = \frac{4k_B T \omega_{NR}}{Qm} \frac{1}{\left(\omega^2 - \omega_{NR}^2\right)^2 + \left(\frac{\omega\omega_{NR}}{Q}\right)^2}$$

where, $\int_0^\infty S_x \frac{d\omega}{2\pi} = \langle x^2 \rangle = \frac{4k_B T}{\omega_{NR}^2 m}$

obeying the equipartition of energy, where $k_B$ is the Boltzmann constant.

The expected total charge noise spectrum is

$$S_Q(\omega) = S_Q^{NR}(\omega) + S_{SSET} = \left(\frac{dC_{NR}}{dx}V_{NR}\right)^2 S_x(\omega) + S_{SSET}$$

where, $S_{SSET}$ is the white, SSET charge noise originating from the cryogenic preamplifier in our setup.

**W**e measure the charge noise of the RFSET detector around the mechanical resonance for temperatures from 30 mK to 550 mK. We find a noise peak at the expected mechanical resonance frequency (identified earlier by driving the nanomechanical resonator), sitting upon a white background, $S_{SSET}$. The charge noise power data accurately fits the expected harmonic oscillator response function. We extract both the background noise power, $S_{SSET}$ and the integral of the resonator noise power which is a measure of the resonator position variance:

$$P_{NR} = \int S_Q^{NR}(\omega)\frac{d\omega}{2\pi} \propto \langle x^2 \rangle = \frac{k_B T}{k}$$

The above equation is true when the backacation effects of the SSET are negligilble. In practice, since we cannot totally decouple the SSET from the resonator, we use the data taken at $V_{NR}=1V$, where the observed back-action is negligible, as temperature calibration curve. We have checked the validity of this calibration by converting the integrated power in units of charge (using the amplitude of the reference 20.5MHz sideband recorded at the time of measurement) and comparing the slope of the charge noise signal against bath temperature with the theoretical value calculated from device parameters determined independently (see section 'sample characteristics'). The frequency, quality factor, and integrated noise power of each spectrum are determined by least-square fitting to a harmonic oscillator response function (see Figure S2).

**Quantum Limit calculations**:

To calculate the minimum uncertainty in resonator displacement that this device can reach, we calculate the displacement noise from two contributions: the shot noise of the SSET current and the displacement noise produced by the backaction force of the SSET onto the resonator. As the calculation involves a number of subtleties, we present a detailed exposition to ensure quality.

*Forward-coupled noise:*

In practice, the position noise, $S_x$, is dominated by the noise floor of the preamplifier used to read out the SSET. Ideally, a measurement based on the same method would be limited by the SSET shot noise. The low-frequency limit ($\omega_{NR} \ll \Gamma, \frac{E_J}{\hbar}$) of the shot noise is given as

$$S_I(\omega) \equiv \int_{-\infty}^{\infty} dt\, e^{i\omega t} \langle I(t)I(0) \rangle$$

Note that we will use this convention for noise spectral densities in all that follows, unless explicitly noted; this convention corresponds to a "two-sided" spectral density. The shot noise and average current are related by the Fano factor, $f = S_I / eI_{DS}$. For a SSET biased near the JQP resonance, f is expected to be *at most* 2, which corresponds to the uncorrelated tunneling of charge 2e Cooper pairs (**6**). We assume this worst-case scenario, and calculate $S_I$ using f=2, and the measured value of $I_{DS}$ at our chosen SET operating point.

This current noise produces a displacement noise given by

$$S_x = \frac{S_I}{(dI_{DS}/dx)^2}.$$

The derivative of the current with respect to the oscillator position is readily calculated from:

$$\frac{dI_{DS}}{dx} = \frac{dI_{DS}}{dV_g}\frac{dV_g}{dQ}\frac{dQ}{dx} = \frac{dI_{DS}}{dV_g}\frac{e}{C_g}\frac{\Delta F}{2E_c}$$

The derivative $dI_{DS}/dV_g$ can be calculated numerically from our data to find $dI_{DS}/dx$=12.5A/m at $V_{NR}$=1V, giving a position noise $S_x^{1/2}$=93am/Hz$^{1/2}$ at our point of bias ($I_{DS}$=0.8nA) and $V_{NR}$=15V.

By comparison, the preamplifier noise floor yields

$$S_x = \frac{1}{2}\frac{S_Q}{\left(V_{NR}\frac{dC_{NR}}{dx}\right)^2} = 175\, am/\sqrt{Hz}$$

in the best case ($S_Q^{1/2}$=10µe/ Hz$^{1/2}$) at $V_{NR}$=15V. During our measurements, the sensitivity was lower ($S_x^{1/2}$=4.5fm/Hz$^{1/2}$ with $S_Q^{1/2}$=170µe/Hz$^{1/2}$ and $V_{NR}$=15V), since the amplitude of the SSET microwave carrier signal had to be kept small to limit the excursion of the bias point.

It is interesting to note that if one does not lower the microwave drive, and we optimize our biases for the best position sensitivity, we find a record value of 3.5 10$^{-16}$ m/Hz$^{1/2}$, shown in Figure S3.

Note that the $S_x$ values quoted here are "two- sided" spectral densities. To make a meaningful comparison with other experimental works, the value quoted above should be multiplied by 2.

*Back-action*

The charges hopping on and off the island of the SSET produce stochastic backaction forces which drive the resonator. For the relatively low frequency resonator used here ($\omega_{NR} \ll \Gamma, \frac{E_J}{\hbar}$), this force noise is frequency-independent and its amplitude is obtained by taking the $\omega \to 0$ limit of equ. 4b:

$$S_F(\omega_{NR}) \approx 2m\gamma_{SSET}k_B T_{SSET}$$

Using our experimental values for $T_{SSET}$ and $\gamma_{SSET}$ we find $S_F^{1/2}$=0.63aN/Hz$^{1/2}$ at $V_{NR}$=1V. By contrast, the theoretical predictions lead to $S_F^{1/2}$=0.2aN/Hz$^{1/2}$. As mentioned above, the excess back-action might be caused by incoherent transport mechanisms which also broaden the current peaks at the JQP resonance.

*Quantum Limit*

The back-action forces produce a displacement noise given by:

$$S_X^{BA}(\omega) = \frac{S_F(\omega)}{k^2}\frac{\omega_{NR}^4}{\left(\omega_{NR}^2 - \omega^2\right)^2 + \left(\omega\gamma_{NR}\right)^2}$$

However, the position noise density is constrained by quantum mechanics and the uncertainty principle (ref. 7 and references therein):

$$\sqrt{S_x S_F} \geq \frac{\hbar}{2}.$$

Using our theory, the theoretical values for our device gives an optimum value of

$$\sqrt{S_x S_F} \geq 4\frac{\hbar}{2}.$$

By contrast, our experimental values for the back-action forces give

$$\sqrt{S_x S_F} = 15\frac{\hbar}{2}.$$

Note that the only theoretical assumption we have used in making this calculation was that the Fano factor determining the shot noise had its "worst-case" value of f=2; all other parameters were experimentally determined. In terms of RMS displacement uncertainty, this value is 3.9 away from the quantum limit at the optimal coupling of $V_{NR}$=0.4V:

$$\frac{\Delta X_{RMS}}{\Delta X_{QL}} = 3.9, \text{ where } \Delta X_{QL} = \sqrt{\frac{\hbar \omega_{NR}}{2k}}.$$

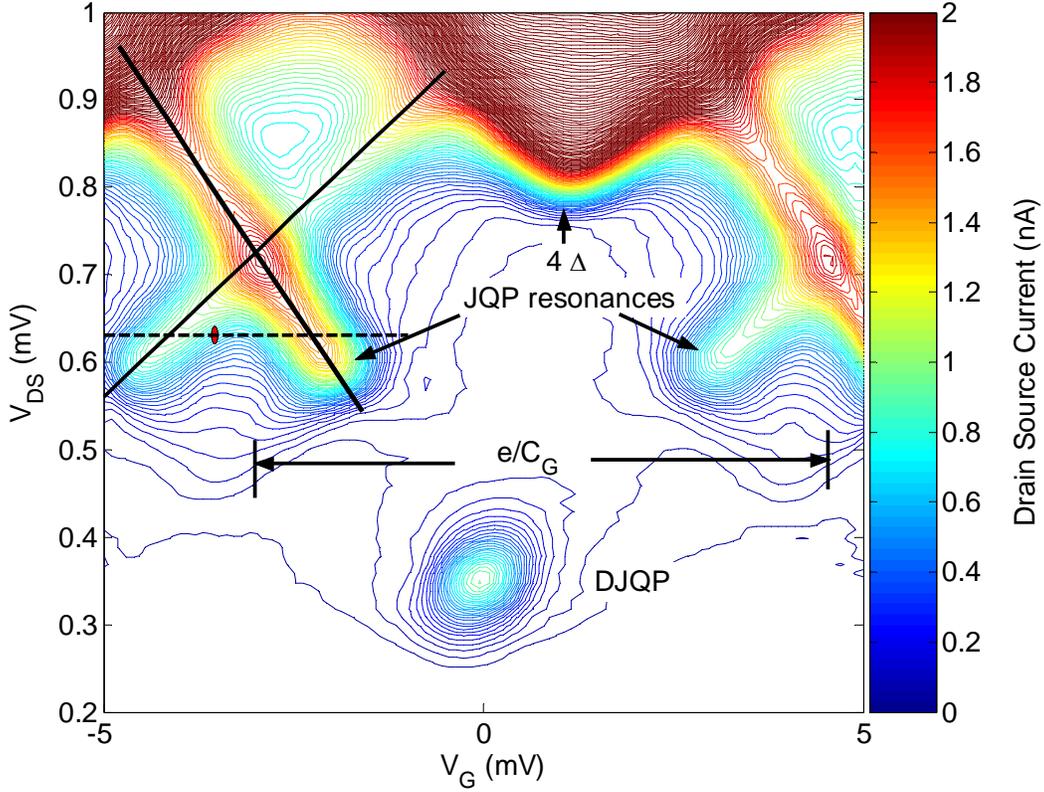

**FIGURE S1**: DC current-voltage characteristics ($I_{DS}$ vs $V_{DS}$ vs $V_{GATE}$) of the SSET. The bias point is shown as red colored ellipse on the left side of the map. The $C_G$ can be determined from the current modulation. The black solid line (negative slope line) along one of the JQP resonance is $-C_G/C_1$ and the slope of the other solid black line is $C_G/(C_\Sigma - C_1)$. These two slopes are used to calculate the junction capacitances. Figure 4 was taken with scans of $V_G$ along dashed horizontal line through both JQP resonances.

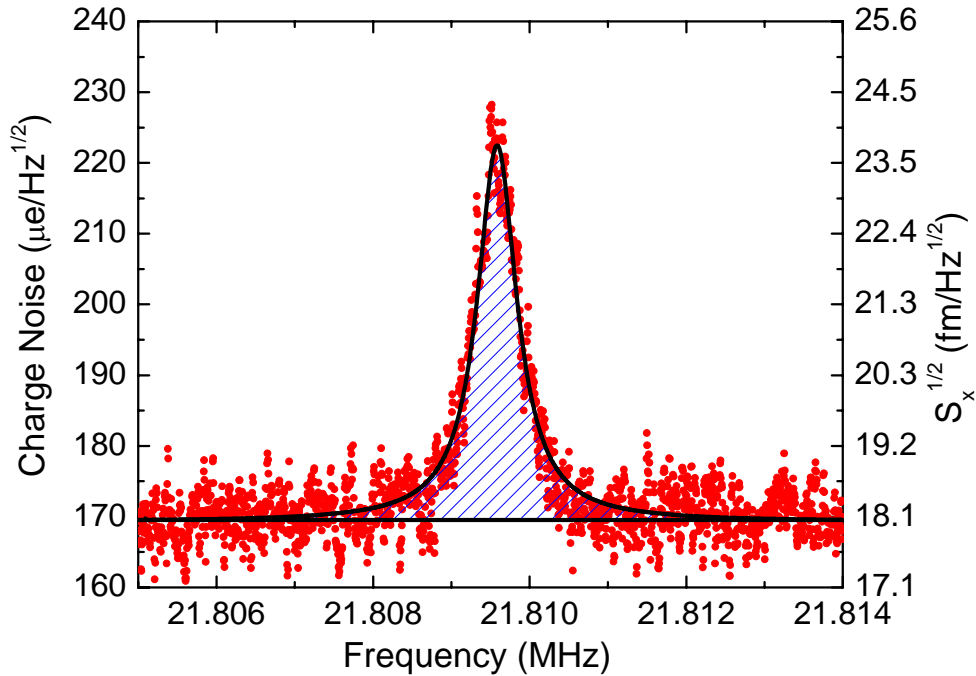

**Figure S2**

**FIGURES2**: Noise spectrum measured at $V_{NR}$=5V and $T_{Bath}$ = 100 mK. The fit (solid black line) is used to extract the resonance frequency, damping coefficient, and integrated power (hatched area). For this spectrum $\omega/2\pi$= 21.809584 MHz+/- 3 Hz, Q = 37,992+/- 716, $<Q^2>$ = 18.8 ·$10^{-6}$ $e^2_{RMS}$ +/- 0.4·$10^{-6}$ $e^2_{RMS}$. [Note that the displacement noise power used in this figure is the "two sided" spectral density.]

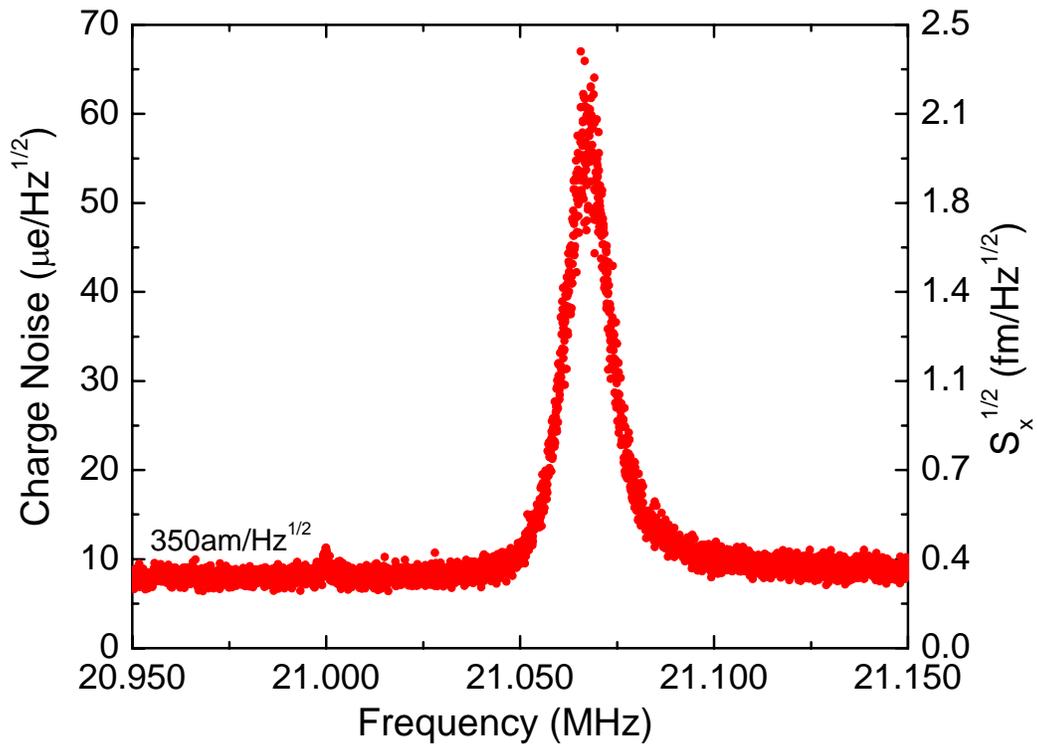

**Figure S3** shows the best position sensitivity that we have been able to achieve with the device. The spectra was taken at $V_{NR}$=15V with the RFSET optimized for maximum gain. [Note that the displacement noise power used in this figure is the "two sided" spectral density.]